\documentstyle[prl,aps,epsfig,floats,twocolumn
]{revtex}

\begin{document}
\draft
\title{Nonperturbative Renormalization Flow and Essential Scaling for the Kosterlitz-Thouless Transition}
\author{G. v. Gersdorff\cite{email1} and C. Wetterich\cite{email2}}
\address{Institut f. Theoretische Physik, Philosophenweg 16, 69120 Heidelberg}
\maketitle
\begin{abstract}
The Kosterlitz-Thouless phase transition is described by the nonperturbative renormalization flow of the two dimensional $\varphi^4$-model. The observation of essential scaling demonstrates that the flow equation incorporates nonperturbative effects which have previously found an alternative description in terms of vortices. The duality between the linear and nonlinear $\sigma$-model gives a unified description of the long distance behaviour for $O(N)$-models in arbitrary dimension $d$. We compute critical exponents in first order in the derivative expansion.
\end{abstract}

\narrowtext
The Kosterlitz-Thouless phase transition \cite{KosThou73} may describe the critical behaviour of various two dimensional systems. 
It poses a challenge to our theoretical understanding due to several uncommon features. 
The low temperature phase exhibits a massless Goldstone boson like excitation despite the fact that the global $U(1)$ symmetry is not spontaneously broken by a standard order parameter. 
In this phase the critical exponents depend on the temperature. 
In the high temperature phase the approach to the transition is not governed by critical exponents but rather by essential scaling. 
Modern nonperturbative renormalization group methods have described the characteristic properties of the low temperature phase qualitatively \cite{GraeWett94}. 
In this letter we extend this analysis to the essential scaling in the high temperature phase and provide for a quantitatively accurate discussion of the low temperature phase. 
We present a universal renormalization group description of models with $O(N)$ symmetry in arbritrary dimension.

We employ the concept of the effective average action $\Gamma_k$ \cite{Wett91},
which equals the effective action $\Gamma$ apart from the fact that in the former only fluctuations with momenta larger than $k$ are included. For large momentum scales $\Gamma_k$ therefore equals some appropriately regularized microscopic action while for $k=0$ we have $\Gamma_0=\Gamma$. Thus $\Gamma_k$ interpolates between  microscopic and  macroscopic scales.
Since $\Gamma$ is the generating functional for the 1PI correlation functions it specifies directly the quantities of interest like the correlation length $\xi=m_R^{-1}$.
The flow of $\Gamma_k$ is obeys an {\em exact} renormalization group equation \cite{Wett91},
\begin{equation}
\partial_t \Gamma_k[\varphi]=\frac{1}{2}\text{Tr}\left((\Gamma_k^{(2)}[\varphi]+R_k)^{-1} \cdot \partial_tR_k  \right).\label{mastergleichung}
\end{equation}
Here $\Gamma^{(2)}_k$ is the second functional derivative, $\partial_t$ denotes the logarithmic derivative $k\cdot\partial/\partial k$ and the trace (in momentum space) reads $\text{Tr}=\sum_{a=1}^N\intop d^dq/(2\pi)^d$.
The cutoff $R_k(q^2)$ suppresses the low momentum modes. 
We use a cutoff of the form
\begin{equation}
R_k(q^2)=Z_k q^2 r(q^2/k^2)=\frac{Z_k q^2}{\exp(q^2/k^2)-1},
\end{equation}
where the wave function renormalization $Z_k$ will be fixed later. 
In order to solve equation (\ref{mastergleichung}) numerically one has to truncate the most general form of $\Gamma_k$.
We introduce dimensionless, renormalized fields, $\tilde\varphi_a=Z_k^{1/2}k^{(2-d)/2}\varphi_a$, $\tilde\rho=(1/2)\tilde\varphi_a\tilde\varphi_a$ and parametrize $\Gamma_k$ in first order in a derivative expansion by
\begin{eqnarray}
\Gamma_k&=&\intop d^dx \left\{k^du_k(\tilde\rho) + \frac{1}{2}k^{d-2}z_k(\tilde\rho)\partial_\mu\tilde\varphi_a\partial_\mu\tilde\varphi_a\nonumber\right.\\
&&\left.+\frac{1}{4}k^{d-2}\tilde y_k(\tilde\rho)\partial_\mu\tilde\rho\partial_\mu\tilde\rho+O(\partial^4)\right\}.\label{truncation}
\end{eqnarray}
The flow of $\Gamma_k$ is then given by the flow of the quantities $u$, $z$ and $\tilde y$ which are dimensionless and depend only on the $O(N)$-invariant $\tilde\rho$ and on $k$.
 We denote by $\kappa$ the running minimum of the potential $u_k(\tilde\rho)$ and fix $Z_k$ by requiring $z_k(\kappa)=1$.

The partial differential equations \cite{Wett91,Wett93,TetWett94}
\begin{eqnarray}
\partial_tu & =& 
	-du+(d-2+\eta)\tilde\rho u'\nonumber\\
	&&+2v_d(N-1)l^d_0(w,z,\eta)+2v_dl_0^d(\tilde w,\tilde z,\eta),
\label{dudt}
\end{eqnarray}

\begin{eqnarray}
\partial_t z
	&=&\eta z +\tilde\rho z'(d-2+\eta)\nonumber\\
	&&-(4v_d/d)\tilde\rho^{-1}\bigl\{\bigr.
	m^d_{2,0}(w,z,\eta)\nonumber\\
	&&-2m^d_{1,1}(w,\tilde w,z,\tilde z,\eta)
	+m^d_{0,2}(\tilde w,\tilde z,\eta)\bigl.\bigr\}\nonumber\\
	&&-2v_d(\tilde z-z)\tilde\rho^{-1}\bigl\{\bigr.l_1^d(\tilde w,\tilde z,\eta)\nonumber\\
	&&-(2/d)(\tilde z-z)l^{d+2}_2(\tilde w,\tilde z,\eta)\bigl.\bigr\}\nonumber\\
	&&-2v_dz'\bigl\{\bigr.(N-1)l_1^d(w,z,\eta)\nonumber\\
	&&-(8/d)n^d_{1,1}(w,\tilde w,z,\tilde z,\eta)\nonumber\\
	&& +\left(5+2z''\tilde\rho/z'\right)
	l_1^d(\tilde w,\tilde z,\eta)\nonumber\\
	&&-(4/d)z'\tilde\rho l_{1,1}^{d+2}(w,\tilde w,z,\tilde z,\eta)\bigl.\bigr\},
\label{dzdt}
\end{eqnarray}

\begin{eqnarray}
\partial_t\tilde z
	&=&\eta\tilde z +\tilde\rho\tilde z'(d-2+\eta)\nonumber\\
&&	-2v_d(\tilde z'+2\tilde\rho\tilde z'')l_1^d(\tilde w, \tilde z,\eta)\nonumber\\
&&	+8v_d\tilde\rho\tilde z'(3u''+2\tilde\rho u''')
	l_2^d(\tilde w, \tilde z,\eta)\nonumber\\
&&	+4v_d\left(2+1/d\right)\tilde\rho(\tilde z')^2
	l_2^{d+2}(\tilde w, \tilde z,\eta)\nonumber\\
&&	-(8/d)v_d\tilde\rho(3u''+2\tilde\rho u''')^2
	\tilde m_4^d(\tilde w, \tilde z,\eta)\nonumber\\
&&	-(16/d)v_d\tilde\rho\tilde z'(3u''+2\tilde\rho u''')
	\tilde m_4^{d+2}(\tilde w, \tilde z,\eta)\nonumber\\
&&	-(8/d)v_d\tilde\rho(\tilde z')^2
	\tilde m_4^{d+4}(\tilde w, \tilde z,\eta)\nonumber\\
&&	+(N-1)v_d\bigl\{\bigr .\nonumber\\
&&	-2\left(\tilde z'-\tilde\rho^{-1}(\tilde z -z)\right)
	l_1^d(w,z,\eta)\nonumber\\
&&	-(8/d)\tilde\rho (u'')^2m_4^d(w,z,\eta)\nonumber\\
&&	-(16/d)\tilde\rho u''z'm_4^{d+2}(w,z,\eta)\nonumber\\
&&	-(8/d)\tilde\rho (z')^2m_4^{d+4}(w,z,\eta)\nonumber\\
&&	+4(\tilde z -z)u''l_2^d(w,z,\eta)\nonumber\\
&&	\bigl .+4\left(z'(\tilde z -z)+
	(1/d)\tilde\rho(z')^2\right)
	l_2^{d+2}(w,z,\eta)\bigr \}	
\label{dztdt}
\end{eqnarray}
invoke the ``threshold functions'' $l_{n_1,n_2}^d$, $m_{n_1,n_2}^d$, $\tilde m_{n_1,n_2}^d$ and $n_{n_1,n_2}^d$ defined by the integral
\begin{equation}
-\frac{1}{2}\intop_0^\infty 
	dyy^{\frac{d}{2}-1}
	\tilde{\partial_t}\left \{ \frac{X}{(p(y)+w)^{n_1}(\tilde p(y)+\tilde w)^{n_2}}\right \},
\end{equation}
with $X=1,\ y (\partial_yp)^2,\ y(\partial_y\tilde p)^2,\ y\partial_y p$ for $l,\ m,\ \tilde m,\ n$ respectively.
We have defined $u'=\partial u/\partial\tilde\rho$ and we use the shorthands
$v_d^{-1}=2^{d+1}\pi^{d/2}\Gamma(d/2),\ 
w=u',\ 
\tilde w=u'+2\tilde\rho u'',\ 
\tilde z=z+\tilde \rho\tilde y,\ 
p(y)=y(z+r(y))$, and   
$\tilde p(y)=y(\tilde z+r(y)).
$
The derivative $\tilde \partial_t$ only acts on the $k$-dependence of the cutoff $R_k$, i.e. 
$\tilde\partial_t p(y)=-y\bigl(\eta r(y)+2y\partial_yr(y)\bigr)$
and we note that $\tilde\partial_t\partial_yp=\partial_y\tilde\partial_tp$. Finally we abbreviate $l_{n,0}^d=l_n^d$ etc., where $l_0^d$ is defined by the rule
$(p+w)^{-n}\rightarrow-\log(p+w)$.
The expression for the anomalous dimension $\eta=-\partial_t\ln Z_k$ can be obtained from the identity $\partial_t z_k(\kappa_k)\equiv 0$. 
For $N=1$ one has $\tilde y=0$, $z=\tilde z$. 

These equations are valid in arbitrary dimension. 
In the very simple approximation $z=\tilde z=1$, $u=\frac{1}{2}\lambda(\tilde\rho-\kappa)^2$ these equations already give a correct qualitative picture for $O(N)$ symmetric models in arbitrary dimension \cite{Wett93}. We show that the present version leads to quantitatively accurate results. This is done by a numerical solution with initial values specified at a microscopic scale $k=\Lambda$.
For $\kappa\gg1$ the evolution is dominated by the $N-1$ Goldstone modes ($N>1$). More precisely, the threshold functions at the minimum vanish rapidly for $\tilde w=2\kappa u''(\kappa)\gg1$. For $N>2$ the coupling of the nonlinear $\sigma$- model for the Goldstone bosons is given by $\kappa^{-1}$.

We concentrate first on $d=2$, where the universality of the $\beta$-function for the nonlinear coupling implies for $\kappa\gg1$ an asymptotic form ($N\geq2$) 
\begin{equation}
\partial_t\kappa=\beta_\kappa=\frac{N-2}{4\pi}+\frac{N-2}{16\pi^2}\kappa^{-1}+{\cal O}(\kappa^{-2}). \label{betakappa0}
\end{equation}
In the linear description one can easily obtain an equation for $\kappa$ by using $\partial_tu_k(\kappa_k)=0$ together with equation (\ref{dudt}). By evaluating the above equations for large $\kappa$ it is possible to compare with (\ref{betakappa0}). 
Previously it was found \cite{Wett93} that in a much simpler truncation one already obtains the correct lowest order in the above expansion.
In order to reproduce the exact two loop result one has, however, to go even beyond the truncation (\ref{truncation}).  

From equation (\ref{betakappa0}) we expect that $\kappa$ will run only marginally at large $\kappa$. As a consequence the flow of the action follows a single trajectory for large $-t$ and can be characterized by a single scale.      	
Notice that the perturbative $\beta$-function vanishes for $N=2$ since the Goldstone bosons are not interacting in the abelian case. Thus for large $\kappa$ one expects a line of fixed points which can be parameterized by $\kappa$. This fact plays a major role in the discussion of the Kosterlitz-Thouless transition below. It is responsible for the temperature dependence of the critical exponents.
We have evaluated $\beta_\kappa=\beta_\kappa^{(1)}(N-2)/(4\pi)+\beta_\kappa^{(2)}\kappa^{-1}(N-2)/(16\pi^2)+...$ numerically from the solution of eq. (\ref{dudt}) - (\ref{dztdt})
and extracted the expansion coefficients for large $\kappa$ (see table \ref{nonabelian}).

\begin{table}[t]
\caption[y]{Nonabelian nonlinear sigma model in $d=2$. We show the ratio between the renormalized mass $m_R$ and the nonperturbative scale $\Lambda_{ERGE}$ in comparison with the known ratio \cite{HasMagNied90} 
invoking $\Lambda_{\overline{MS}}$: $C_{ERGE}=m_R/\Lambda_{ERGE}$, $C_{\overline{MS}}=m_R/\Lambda_{\overline{MS}}$, $C_s=m_R/k_s$. We also display the expansion coefficients for the beta function.}
\label{nonabelian}
\begin{tabular}{clllll}
N	& $C_{ERGE}$ & $C_{\overline{MS}}$& $C_s$&$\beta_\kappa^{(1)}$&$\beta_\kappa^{(2)}$ \\
\hline
3	&2.81$\pm$0.30	&2.94	&1.00	&1.00	&0.79\\
9	&1.22$\pm$0.03	&1.25	&1.05	&1.00	&0.84\\
100	&1.08$\pm$0.04	&1.02	&1.06	&1.00	&0.87\\
\end{tabular}
\end{table}
\begin{figure}[b]
\centering\epsfig{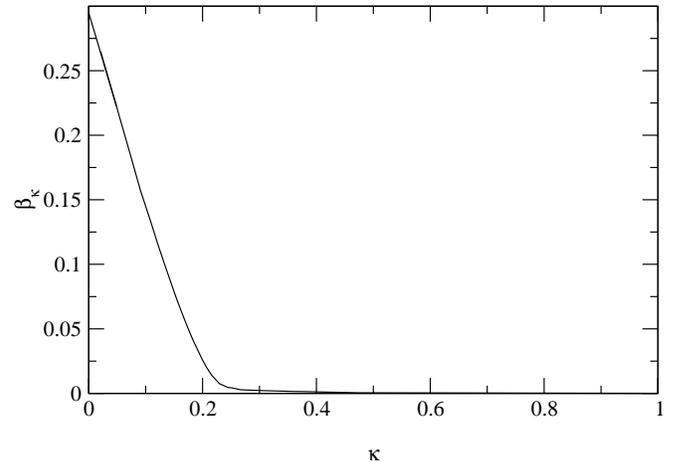}
\caption{The beta function for $d=2,\ N=2$.}
\label{n2betakappa}
\end{figure}
For the nonabelian nonlinear sigma model in $d=2$, $N>2$ there exists an exact expression \cite{HasMagNied90} for the ratio of the renormalized mass $m_R$ and the scale $\Lambda_{\overline{MS}}$ which characterizes the two loop running coupling in the $\overline{MS}$ scheme by dimensional transmutation.
 The flow equation (\ref{mastergleichung}) together with a choice of the cutoff $R_k$ and the initial conditions also defines a renormalization scheme. The corresponding parameter $\Lambda_{ERGE}$ specifies the two loop perturbative value of the running coupling $\kappa^{-1}$ similar to $\Lambda_{\overline{MS}}$ in the $\overline{MS}$ scheme. The numerical solution of the flow equation permits us to compute $m_R/\Lambda_{ERGE}$. (Two loop accuracy would be needed for a quantitative determination of $\Lambda_{ERGE}/\Lambda_{\overline{MS}}$.) In table \ref{nonabelian} we compare our results with the exact value of $m_R/\Lambda_{\overline{MS}}$. We also report the ratio $m_R/k_s$ with $k_s$ defined by $\kappa(k_s)=0$.

\begin{figure}[t]
\centering\epsfig{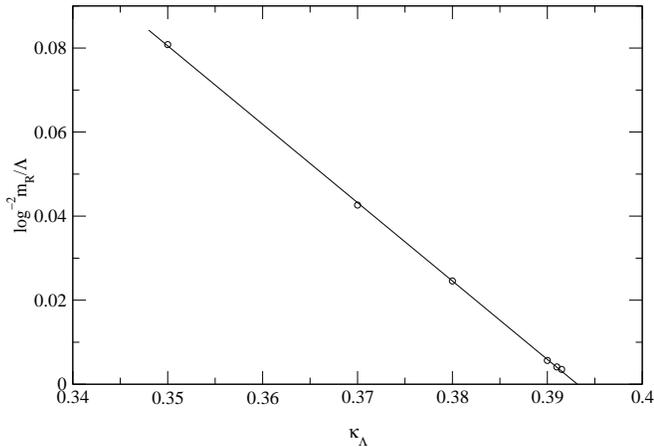}
\caption{Essential scaling for $d=2,\ N=2$. The renormalized mass $m_R$ is plotted as a function of $\kappa_\Lambda=\kappa_{\Lambda*}-H(T-T_c)$.}
\label{escfig}
\end{figure}
\begin{figure}[b]
\centering\epsfig{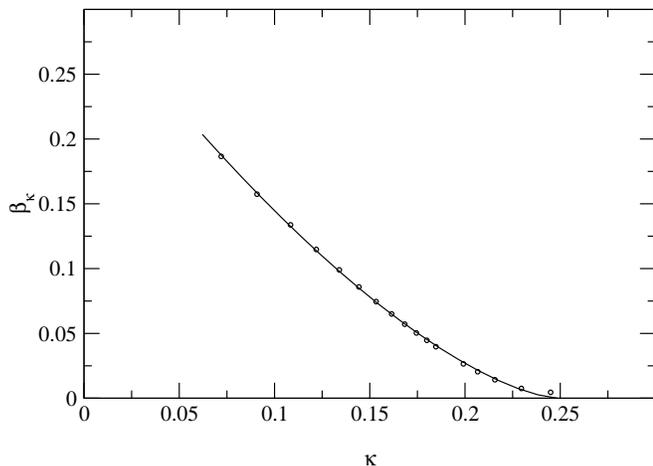}
\caption{Details of the beta function for $d=2$, $N=2$. The curve is a fit to eq. (\ref{betakappa param}).}
\label{esc2}
\end{figure}
The abelian case, $N=2$, is known to exhibit a special kind of phase transition which is usually described in terms of vortices \cite{KosThou73}. 
The characteristics of this transition are a massive high temperature phase and a low temperature phase with divergent correlation length but zero magnetization. The anomalous dimension $\eta$ depends on $T$ below $T_c$ and is zero above. It takes the exact value $\eta_*=0.25$ at the transition. The most distinguishing feature is essential scaling for the temperature dependence of $m_R$ just above $T_c$,
\begin{equation}
m_R\sim e^{-\frac{b}{(T-T_c)^\zeta}},\quad \zeta=\frac{1}{2}\label{esc}.
\end{equation}
\begin{figure}[t]
\centering\epsfig{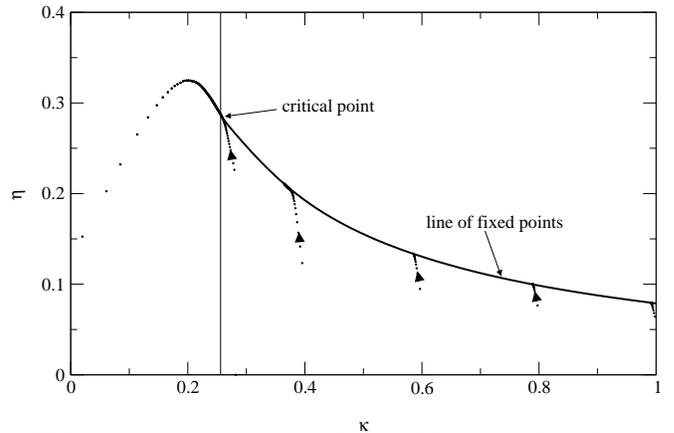}
\caption{Temperature dependence of the anomalous dimension $\eta$ for the low temperature phase, $d=2$, $N=2$. The line of fixed points is characterized by $\kappa$ and ends in the critical point for the Kosterlitz-Thouless phase transition. We also show the flow towards the line of fixed points and the flow in the high temperature phase away from the critical point (left). The spacing between the points indicates the speed of the flow.}
\label{etakappa}
\end{figure}Our 
aim is to draw a picture of this transition from the flow equations without ever examinig vortex configurations.

We have already mentioned the existence of a line of fixed points for large values of $\kappa$, which is relevant for the low temperature phase. The contribution of a massless (Goldstone) boson in the renormalization group equation ($w(k)=0$) is 
responsible for the finite value of $\eta$. This in turn drives the expectation value of the unrenormalized field to zero (even for a nonvanishing renormalized expectation value $\kappa$), 
\begin{equation}
\rho_0=\kappa/Z_k\sim\kappa(k/\Lambda)^\eta.
\end{equation}  
We can observe this line of fixed points to a good approximation (cf.\ fig.\ \ref{n2betakappa}), although the vanishing of $\beta_\kappa$ is not exact (we find $\beta_\kappa\approx -4\cdot10^{-5}\kappa^{-1}+{\cal O}(\kappa^{-2})$). 
The line of fixed points ends at a phase transition which corresponds to a microscopic parameter $\kappa_{\Lambda*}$ as an initial value of the flow. 
In order to verify essential scaling we have to examine the flow for values of $\kappa_\Lambda$ just below that point, $\kappa_\Lambda=\kappa_{\Lambda*}+\delta\kappa_\Lambda$, $\delta\kappa_\Lambda\sim-(T-T_c)$. Then $\kappa_k$ crosses zero at the scale $k_s$ and we find the mass by continuing the flow in the symmetric regime (minimum at $\kappa=0$). In figure \ref{escfig} we plot $(\ln{(m_R/\Lambda}))^{-2}$ against $\kappa_\Lambda$ and find excellent agreement with the straight line (\ref{esc}).

How does $\beta_\kappa$ have to look like in order to yield essential scaling? Since there is only one independent scale near the transition, one expects $m_R(T)=C_sk_s(T)$, where $k_s$ denotes the scale at which $\kappa$ vanishes, i.e. $\kappa(k_s,T)=0$. 
For $\kappa$ close to and below $\kappa_*$ we parameterize $\beta_\kappa$ (this approximation is not valid for very small $\kappa$)
\begin{equation}
\beta_\kappa=\frac{1}{\nu}\cdot(\kappa_*-\kappa)^{\zeta+1}.\label{betakappa param}
\end{equation}
For conventional scaling one expects $\zeta=0$ and the correlation length exponent is given by $\nu$. Integrating equation (\ref{betakappa param}) yields for $\zeta\neq0$, $\delta\kappa=\kappa-\kappa_*$:
\begin{equation}
\ln (k/\Lambda)=\frac{\nu}{\zeta}\left(\frac{1}{(-\delta\kappa)^\zeta}-\frac{1}{(-\delta\kappa_{\Lambda})^\zeta}\right)\label{integriert}.
\end{equation}
For $k=k_s$ the first term $\sim (-\delta\kappa)^{-\zeta}$ is small and independent of $T$ (since $-\delta\kappa(k_s)=\kappa_*$) and equation (\ref{integriert}) yields the essential scaling relation (\ref{esc}) for $\zeta=1/2$.
Usually, the microscopic theory is such that one does not start immediately in the vicinity of the critical point and the approximation (\ref{betakappa param}) is not valid for $k \approx \Lambda$. 
However, if one is near the critical temperature the trajectories will stay close to the critical one, $\kappa_c(t)$, with $\kappa_c(0)=\kappa_{\Lambda*}$. 
This critical trajectory converges rapidly to its asymptotic value $\kappa_*$ and $\beta_\kappa$ gets close to eq. (\ref{betakappa param}) at some scale $\Lambda'<\Lambda$. As a result, one may use equation (\ref{integriert}) only in its range of validity ($k<\Lambda'$) and observe that $\delta\kappa_{\Lambda'}$ is also proportional to $T_c-T$.
The numerical verification of (\ref{betakappa param}) is quite satisfactory: Fitting our data  yields $\kappa_*=0.248$, $\zeta=0.502$ and $\nu^{-1}=2.54$. The uncertainty for $\zeta$ is approximately $\pm 0.05$.
The numerical values of $\beta_\kappa$ and the approximation (\ref{betakappa param}) are shown in figure \ref{esc2}.

\begin{table}[t]
\caption{Critical exponents $\nu$ and $\eta$ for $d=2$. We compare each value with the exact result.}
\label{exp2}
\begin{tabular}{cllll}
$N$
&\multicolumn{2}{c}{$\nu$}
&\multicolumn{2}{c}{$\eta$}
\\
\hline
0	&0.70	&0.75	&0.222	&0.2083...	\\
1	&0.92	&1	&0.295	&0.25\\
2	&--	&--	&0.287	&0.25
\end{tabular}
\end{table}
\begin{table}[t]
\caption[y]{Critical exponents $\nu$ and $\eta$ for $d=3$ (see \cite{SeideWett99} for $N=1$). For comparison we list in the third and fifth column an ``average value'' from various other methods \cite{Zinn99}.}
\label{exp3}
\begin{tabular}{cllll}
$N$
&\multicolumn{2}{c}{$\nu$}
&\multicolumn{2}{c}{$\eta$}
\\
\hline
0	&0.590	&0.5878	&0.039	&0.292	\\
1	&0.6307	&0.6308	&0.0467	&0.0356	\\
2	&0.666	&0.6714	&0.049	&0.0385	\\
3	&0.704	&0.7102	&0.049	&0.0380	\\
4	&0.739	&0.7474	&0.047	&0.0363	\\  
10	&0.881	&0.886	&0.028	&0.025	\\
100	&0.990	&0.989	&0.0030	&0.003	\\
\end{tabular}
\end{table}
One can use the information from figure \ref{escfig} or \ref{esc2} in order to determine $\kappa_*$ and therefore $\eta_*=\eta(\kappa_*)$, the anomalous dimension at the transition. We plot $\eta$ against $\kappa$ in figure \ref{etakappa}. One reads off $\eta_*=0.287\pm0.007$ where the error reflects the two methods used to compute $\kappa_*$ and does not include the truncation error.
For $\kappa_\Lambda>\kappa_{\Lambda*}$ or $T<T_c$ the running of $\kappa$ essentially stops after a short ``initial running'' towards the line of fixed points. One can infer from figure \ref{etakappa} the temperature dependence of the critical exponent $\eta$ for the low temperature phase.
In summary all the relevant characteristic features of the Kosterlitz-Thouless transition are visible within our approach. 

We end this letter by reporting the values of the critical exponents obtained in our approximation (\ref{dudt}) - (\ref{dztdt}) for the ``standard'' second order phase transitions for $d=2$, $N=0,1$ and $d=3$, $N\geq0$. In tables \ref{exp2}, \ref{exp3} they are compared with exact results or ``averages'' (only for the simplicity of the display!) of results from various other methods \cite{Zinn99}.
\begin{table}[t]
\caption{Couplings for the scaling solution for $d=2$ and $N=0,1$.}
\label{couplings2}
\begin{tabular}{clllll}
N&$\kappa_*$&$\lambda_*$&$u_{3*}$&$z'_*(\kappa_*)$\\
\hline
0 &0.151	&5.33	&61.6	&$-$0.085				\\
1 &0.265	&5.88	&65.4	&0.868				\\
\end{tabular}
\end{table}
\begin{table}[t]
\caption{Couplings for the scaling solution for $d=3$ and various $N$.}
\label{couplings3}
\begin{tabular}{cllllll}
N&$\kappa_*$&$\lambda_*$&$u_{3*}$&$z'_*(\kappa_*)$&$\tilde z_*(\kappa_*)$\\
\hline
0 &0.03009	&7.399	&78.84	&0.192		&--		\\
2 &0.05984	&6.769	&51.25	&$-$0.0415	&1.0602		\\
3 &0.07651	&6.256	&39.46 	&$-$0.0920	&1.0695		\\
4 &0.09414 	&5.752	&30.52	&$-$0.1107	&1.0789		\\
10&0.2162	&3.365	&8.17	&$-$0.0584	&1.1144		\\
100&2.2313	&0.3779	&0.0947	&$-$0.000759	&1.1468		\\
\end{tabular}
\end{table}The 
agreement is very satisfactory! We also characterize in tables \ref{couplings2}, \ref{couplings3} the scaling solution relevant for the second order transition by quoting $\kappa_*,\ \lambda_*=u''_{*}(\kappa_*),\ u_{3*}=u_*'''(\kappa_*)$ as well as $z_*'(\kappa_*)$ and $\tilde z_*(\kappa_*)$. We conclude that the first order in the derivative expansion of the exact flow equation for the effective average action gives a quantitatively accurate picture of all phase transitions of scalar models in the $O(N)$ universality class for arbitrary dimension $2\leq d\leq4$.





%
%

%
%

\end{document}